*Model Reduction via Parametrized Locally Invariant Manifolds: Some Examples*

Aarti Sawant and Amit Acharya[*]

Civil and Environmental Engineering

Carnegie Mellon University, Pittsburgh, PA 15213, U.S.A.

**Abstract**

A method for model reduction in nonlinear ODE systems is demonstrated through computational examples. The method does not require an implicit separation of time-scales in the fine dynamics to be effective. From the computational standpoint, the method has the potential of serving as a subgrid modeling tool. From the physical standpoint, it provides a model for interpreting and describing history dependence in coarse-grained response of an autonomous system.

## 1. Introduction

We demonstrate a technique for model reduction, or reduction in degrees of freedom, in nonlinear, autonomous systems of Ordinary Differential Equations (ODE). Our primary motivation comes from a desire to develop a systematic conceptual and practical method for multiscale modeling in solid mechanics. However, the ideas we demonstrate are general in scope. The basic premise of our approach is spelled out in Acharya (2005). Here we implement those ideas in the context of three model problems as a first 'proof of principle'. In the rest of this section we review work in the literature most closely related to our approach to set our work in context. This is not meant to be a review on computational model reduction based on global-in-time empirical eigenvector approaches, coarse-graining, or homogenization techniques.

Muncaster (1983), generalizing his work on the kinetic theory with Truesdell (Truesdell and Muncaster, 1980), proposed a formal methodology for deriving coarse theories from corresponding autonomous fine ones. Our work began from trying to utilize these ideas to develop a practically useful and reliable method for coarse-graining and hence we discuss it here. The concept in Muncaster (1983) is quite attractive as it appears to allow an arbitrary physically motivated choice of coarse variables (reduction), and seems to deliver, after some work, a closed coarse theory. However, the hypotheses assumed are geared towards producing an autonomous coarse theory - "identify each coarse state with exactly one solution (of the fine theory)" (Muncaster, 1983). This hypothesis, in effect, disallows a general choice of coarse variables (reduction) in the sense that doing so can produce a severely restricted coarse model, as is easy to convince oneself with some simple examples. Apart from this feature, the equation defining the main ingredient of the proposal in Muncaster (1983) -the gross determiner- is not uniformly well-defined even in the setting of the fine system being a system of ODEs with a subset of its degrees of freedom being the coarse variables (Acharya, 2005). Finally, issues of (non)uniqueness of the gross determiner and the consequent invariance of the developed coarse theory with respect to this non-uniqueness, while recognized, are not addressed. Nevertheless, to our knowledge, it is Muncaster's work that achieves the first systematic abstraction of classical methods like the Kinetic Theory, Center Manifold theory, and Perturbation theory for Invariant Surfaces (Sacker, 1965) to yield a general methodology for model reduction in nonlinear systems.

---

[*] Corresponding author - Tel. (412) 268 4566; Fax. (412) 268 7813; email: acharyaamit@cmu.edu

The restriction in the arbitrariness of the number of required coarse variables to develop an autonomous coarse theory is dealt with rigorously in Inertial Manifold Theory (Foias *et al*., 1988 a) for dissipative fine systems, along with existence results for such manifolds. In the form of its application most closely related to our work, the desire to generate a single-valued coarse-to-fine map restricts the choice of the number and type of coarse variables. This may be intuitively understood in the context of the Lorenz system (Lorenz, 1963); here the fine system is 3-dimensional, and trajectories tend to converge on a topologically complicated, but bounded, set that can be enveloped within another bounded and contiguous region of 3-d phase space. However, it is not possible to represent the enveloping region as a graph of a *single* function over the space consisting of any two of the system degrees of freedom that one might wish to choose as coarse variables. Consequently, the minimum number of coarse degrees of freedom of the type preferred above would seem to be three, and no reduction can be achieved. Moreover, for systems where the dimension, suitably defined, of the eventual set in which almost all trajectories settle down is large, e.g. Navier Stokes equation, the Inertial Manifold is necessarily of large dimension.

To the extent that we can understand the conceptual idea behind the Method of Invariant Manifolds (Gorban and Karlin, 1992) and its implementation as the Method of Invariant Grids (Gorban *et al.*, 2004), it appears that the goal is again to extract a single manifold of 'slow' motions corresponding to dissipative fine dynamics. Coarse variables are restricted by the notion of Thermodynamic Projectors; however, these restrictions do not seem to deliver a lower bound on the number of coarse variables that may be required to parametrize the sought after single manifold, as in Inertial Manifold Theory. Consequently, how such a method might succeed in providing a reduction for a model like the Lorenz system is not clear to us.

The work of Roberts (2003) is notable from the standpoint of creatively implementing Center Manifold Theory to dissipative fluid dynamics including applications where the Theory does not strictly apply in its rigorous form. Here too, the idea is to seek a single slow manifold.

There are a host of methods that have been proposed for computing (un)stable invariant manifolds of dynamical systems (Foias *et al*., 1988 b; Dieci and Lorenz, 1995; Dellnitz and Hohmann, 1997; Krauskopf and Osinga 2004; Guckenheimer and Vladimirsky, 2003) with application to the understanding of dynamics and bifurcations. Such methods would seem to be useful for developing a closed dynamical system for the evolution of a reduced set of variables corresponding to states on these manifolds, but to our knowledge implementations along these lines have not been developed as yet.

The essence of our method is the following: we would like to make a somewhat mathematically arbitrary choice of coarse variables motivated only by physical considerations and the kind of computational power we wish to bring to real-time computation of the evolution of coarse quantities. Our coarse-graining goal is modest in the sense that we make an implicit assumption about the trajectories or trajectory-segments that we are interested in coarse-graining by considering only those that exist within a fixed region, $R$, of fine phase space. We then choose a physically motivated fine-to-coarse projection map that defines a certain number of coarse variables. Typically, the number of coarse variables is less than the dimension of $R$. Next we try to *populate $R$ by many* locally invariant manifolds of dimension equal to the number of coarse variables and explicitly parametrized by them (Acharya, 2005). Here, by *a locally invariant manifold we* mean *a set of points (a bounded hypersurface) with the property that the vector field defining the fine dynamical system is tangent to the set at all of its points*. Also, we obtain each parametrization by solving a PDE for it. With the fine-to-coarse reduction map and the coarse-to-fine parametrizations in hand, it becomes possible to define a closed evolution equation for the coarse variables that is consistent with the fine autonomous system in a definite sense. Because of the use of a set of locally invariant manifolds of dimension less than that of $R$, the coarse evolution is not necessarily that of an autonomous system and displays history dependence (self-intersection of trajectories). This is considered as an explanation of the emergence of physically observed history-dependent behavior from underlying current-state-dependent behavior, that arises as a direct consequence of reduction or coarser observation of the fine system. In the following sections, we strive to establish and illustrate this idea.

We have chosen to refer to the methodology we develop as the method of Parametrized Locally Invariant Manifolds (PLIM).

## 2. The Coarse-Graining Scheme

In this section we briefly review the proposal of Acharya (2005), extending the ideas of Muncaster (1983). The extensions relate to

1. Demonstrating formal uniqueness ( see point c. at the end of this section) and consistency properties of the coarse theory. These properties are crucial for practically useful model reduction in the presence of a somewhat arbitrary choice of coarse variables.

2. Exploring the obstruction to the construction of locally invariant manifolds and the natural connection of the origin of such an obstruction to the idea of reduction in degrees of freedom.
3. Establishing the possibility of consistent model reduction *without* a constraint on the minimum number of coarse degrees of freedom, with the associated price being a non-autonomous coarse dynamics. This idea makes contact with a result of Inertial Manifold Theory that establishes a lower bound on the dimension of the *necessarily autonomous* coarse dynamics that is sought there.

Let states of the fine theory be represented by the symbol $f$ and the collection of all possible fine states be represented by $\Phi$, i.e. $\Phi$ is the phase space of the autonomous fine theory given by

$$\dot{f}(t) = H(f(t)) \; ; \; J(f(t)) = 0 \; ; \; B(f(t)) = 0, \tag{1}$$

where $H$ is some mapping (in general a nonlinear operator) on $\Phi$, $J$ is an operator defining jump conditions, and $B$ is an operator defining boundary conditions. We assume that appending an initial condition endows a nominal sense of uniqueness of solutions, i.e. closure, to the system (1).

Similarly, let 'coarse' states be represented by the symbol $c$, and the collection of all possible coarse states, i.e. the phase space of the to-be-defined coarse theory, by the symbol $\varphi$. A primary ingredient of our procedure is that a function

$$\Pi : \Phi \to \varphi \tag{2}$$

has to be defined based on physical considerations that provides a recipe for generating coarse states from fine states. For example, if the fine variables are time-varying functions of space, the coarse variables may be time varying fields defined as running averages over specified length-scales (i.e. spatial resolution of the coarse theory) of the fine fields. It is of the essence of coarse-graining that in almost all circumstances such a relationship is not invertible in the sense that many fine states will in general correspond to a given coarse state. We assume that the topological structure of $\varphi$ allows, at least, differentiation at each of its elements.

The next ingredient is a general idea of a possible set of initial conditions of the fine theory, solutions emanating from elements of which one is interested in coarse graining. We call this set $I$ and clearly $I \subset \Phi$. The larger this set is (one limit is $\Phi$ itself) the more is the work to be done in setting up the coarse theory.

The main task of the coarse graining procedure is defined as seeking a mapping

$$G : \varphi \to \Phi \tag{3}$$

that satisfies

$$\left.\begin{array}{l} DG(c)\big[D\Pi\big(G(c)\big)\big[H\big(G(c)\big)\big]\big] = H\big(G(c)\big) \\ J\big(G(c)\big) = 0 \,;\, B\big(G(c)\big) = 0 \,;\, \Pi\big(G(c)\big) = c \end{array}\right\} \quad \forall \;\; c \in \varphi, \tag{4}$$

where the symbol $D$ represents a derivative of its argument function. For the present, we assume that a solution exists to (4).

Suppose, now, that we are interested in obtaining a closed theory one of whose solutions represents the coarse dynamics corresponding to a fine solution emanating from the initial condition $f_* \in I$. Let us assume that there exists a $c_* \in \varphi$ such that

$$G(c_*) = f_*. \tag{5}$$

Because of $(4)_4$, this implies that $c_* = \Pi(f_*)$, and consequently an efficient way to check for (5) would be to evaluate $G\big(\Pi(f_*)\big)$ and see whether the result equals $f_*$ or not. Granted (5), we claim that

$$\dot{c}(t) = D\Pi\big(G(c(t))\big)\big[H\big(G(c(t))\big)\big] \,;\, c(0) = c_* \tag{6}$$

is the required *closed* coarse theory. This is so because the fine trajectory $\varGamma$ defined by

$$\varGamma(t) = G\big(c(t)\big) \tag{7}$$

satisfies

$$\begin{aligned} \dot{\varGamma}(t) &= DG\big(c(t)\big)\big[\dot{c}(t)\big] \\ &= DG\big(c(t)\big)\big[D\Pi\big(G(c(t))\big)\big[H\big(G(c(t))\big)\big]\big] = H\big(G(c(t))\big) = H\big(\varGamma(t)\big) \\ J\big(\varGamma(t)\big) &= 0 \,;\, B\big(\varGamma(t)\big) = 0 \,;\, \varGamma(0) = G\big(c(0)\big) = G(c_*) = f_*, \end{aligned} \tag{8}$$

by (6), $(4)_{1,2,3}$, and (5). Consequently, uniqueness of solutions to (1) with the initial condition $f(0) = f_*$ implies that $\varGamma$ is the fine solution whose coarse graining we are interested in. This, along with the fact that $G$ satisfies $(4)_4$, implies that the solution of (6) is indeed the coarse representation of *the* fine solution of interest, i.e. the one issuing from the fine initial condition $f_*$.

It bears emphasis that once a solution $G$ of (4) is obtained and (5) is satisfied for a given fine initial condition $f_*$, then (6) can be solved independently of solving the fine theory exactly or approximately. It should also be noted that this scheme is a reduction of work as (4), even though a difficult, non-standard problem to solve in most circumstances, is independent of time and that the range of $G$ may be expected to include many points belonging to the possible set of initial conditions $I$ so that for these initial conditions $G$ is already available for use in the coarse evolution and does not have to be recomputed.

As shown in Acharya (2005), we note that the definition of the coarse theory as in (6) has the desirable property that if $S$ is some function (e.g. energy, entropy) defined on $\varPhi$ with $\dot{S}$ having a special property along fine trajectories (vanishes, monotonically increases/decreases), then the corresponding function $S^*$ on $\varphi$ also has the same property along coarse trajectories, i.e.

$$S^*(c(t)) = S(G(c(t)))$$
$$\dot{S}^*(c(t)) = DS(G(c(t)))[DG(c(t))[\dot{c}(t)]] = DS(G(c(t)))[H(G(c(t)))] = \dot{S}(G(c(t))). \quad (9)$$

From the physical standpoint, however, the above method of defining, say energy $(S^*)$, in the coarse model may not be optimal – it involves $G$ that is not an 'observable' of the coarse system. As implications of this idea, let us first consider the case where each coarse variable is a field corresponding to a running spatial average of a fine field so that $\varphi \subset \Phi$. In such a case, a physically natural definition of $S^*$ may very well be

$$S^*(c) := S(c), \quad c \in \varphi, \quad (10)$$

in which case $\dot{S}^*(c(t)) \neq \dot{S}(G(c(t)))$, in general, along any coarse trajectory, and one might expect dissipative effects in a coarse theory corresponding to an energy-conserving fine theory. As another example, consider a collection of coupled nonlinear oscillators forming an unforced, undamped, fine system according to classical mechanics. A natural choice of coarse variables, $c$, might be the positions and momenta of a subset of the oscillators. Now define energy for the coarse theory as the sum of the kinetic and potential energies of the retained oscillators, say $\tilde{S}(c(t))$. It is clear from the demonstration (9) that $\dot{\tilde{S}}(c(t)) \neq \dot{S}(G(c(t)))$, when the coarse evolution is defined by our method. We think of this argument as a simple device for interpreting the emergence of non-conservative behavior from conservative fine dynamics. Of course, this does not prove dissipation in coarse response; however, we believe that with a physical choice of coarse variables the emergence of dissipation can also be shown.

Before moving on to examples demonstrating the ideas above, we list some essential issues that have to be dealt with in order to have an implementable algorithm. Due to the generality of the possible applications but the simplicity of the main idea, we lay out the algorithm in the context of a specific example in Section 3.1.

a. Essentially, computing the $G$ function amounts to computing locally invariant manifolds in fine phase space. For the drastic reductions that we have in mind dictated by the choice of coarse variables, both in number and type, the global invariant manifolds containing trajectories of interest are expected to 'fold', when viewed as hypersurfaces over coarse space. Moreover, associating a single manifold with each coarse state is not, in general, adequate for a drastic reduction and eliminates the prediction of physically observed history dependent effects. For all these reasons, we divide coarse phase space into local blocks and precompute and store a collection of $G$ s over each *local* block of coarse phase space.

b.  In reality, only a certain set of fine initial conditions will belong to the ranges of the computed $G$ s representing lower dimensional manifolds. Suppose in the process of coarse evolution a coarse state is attained that corresponds to a fine state at the boundary of a computed manifold. The function $H$ may be used at such a fine state to obtain an estimate of the 'next' fine state to be utilized in coarse evolution. However, such a fine state may not be one that belongs, or is close, to the range of precomputed $G$ s. In such a case, the computation needs to be stopped or a supplemental $G$ computed and added to the collection before proceeding. Such supplemental computations result in enhancement of the range of applicability of the theory.

c.  The solution for $G$ is not unique; since it enters into the definition of the coarse theory, is coarse evolution affected by the choice? It can be shown (Acharya, 2005) that the choice does not matter as long as the initial condition of the fine trajectory in being coarse-grained belongs to the range of the distinct $G$ candidates under consideration for use in the coarse theory.

d.  At fine states where

$$D\Pi(f)[H(f)] = 0 \text{ with } H(f) \neq 0 \tag{11}$$

the PDE for the determination of $G$ is not well-defined. Moreover, such states cannot be ruled out as belonging to the range of $G$ *a priori*. If such a state is encountered in the course of coarse evolution, one perturbs the problematic, attained coarsely-defined fine state by the fine vector field at that state, i.e. by $\varepsilon H(G(c))$ with $0 < \varepsilon \ll 1$, to define a new state where, presumably, (11) is not satisfied. This requires $H(G(c))$ to not be tangent to the zero set of the, generally, vector valued function $D\Pi(\cdot)[H(\cdot)]$. In case it is, and *if* the choice of coordinates for the parametrization is held to be immutable, then guidance for closed, coarse evolution out of such states cannot be obtained from the structure of the fine dynamics. This issue has interesting implications related to the connection between coarse and fine equilibria and when a correspondence, or lack of it, between them may be acceptable depending upon the physical choices of $\Pi$, but we do not delve into the matter in this paper.

e.  $G$ functions with real valued components may not be defined in all coarse space of interest. The regions of coarse space where this might occur may not be known *a priori*. Seeking complex-valued solutions in coarse domain and then ignoring points where all components of $G$ are not real-valued alleviates the problem. If $c$ is a coarse state in the vicinity of the boundary of such a region, a move to a different manifold, dictated essentially by $H(G(c))$, is required. We illustrate this idea in the context of the reduction of the system representing a simple linear oscillator.

## 3. Model Problems

In this section we illustrate through examples that the method can be implemented successfully. The examples also provide insight into the nature of the coarse theory (despite appearances, its solutions are necessarily history-dependent) and important issues that have to be resolved to construct the family of locally invariant manifolds (family of $G$ s) to accommodate a reasonable set of initial conditions. The first three examples belong to the category where the coarse variables are simply a subset of the fine degrees-of-freedom and coarse equilibria not in correspondence with fine equilibria are not acceptable. We consider this situation as a stringent test of methodology as the coarse problem has to essentially produce/approximate actual fine solutions. We intentionally choose a dissipative (Lorenz system) and a Hamiltonian system (Chorin *et al*., 2002). For the dissipative system, we parametrize a region around the attractor; for the Hamiltonian system, we parametrize an arbitrary region of phase space with the understanding that the coarse model would need extension through further computation (with the same methodology) for coarse graining trajectories/sections of trajectories lying outside the parametrized region. The fourth example demonstrates our ideas of dealing with PDE systems in the context of 1-d heterogeneous elastodynamics. We demonstrate reasonable success for a preliminary implementation and also suggest avenues for further improvement.

Finally, it should be noted that the goal of computation of the collection of locally invariant manifolds for coarse-graining is not to generate exact solutions to the fine dynamical system or the computation of globally invariant manifolds, e.g. (un)stable manifolds of fixed points. Instead, the goal is simply to formulate the dynamics of retained variables in a self-consistent approximation without ad-hoc, and possibly incorrect, physical assumptions. Consequently, there exists some latitude in the quality of the approximations computed. The geometric objects we construct are closer to foliations/laminations with locally invariant leaves with the exception that these 'leaves' may intersect.

### 3.1. Lorenz System

The fine set of equations is given as (Lorenz, 1963)

$$\dot{x} = \sigma(y-x);$$
$$\dot{y} = rx - y - xz;$$
$$\dot{z} = xy - bz;$$
$$\sigma = 10, b = 8/3, r = 25$$
(12)

For $r > 1$ the system has three fixed points ($x = y = z = 0$; $x = y = \pm\sqrt{b(r-1)}$, $z = r-1$), as shown by Lorenz (1963). For the assumed values of $\sigma$ and $b$, Lorenz (1963) also shows that the three fixed points become hyperbolic saddles as $r$ crosses the linear stability threshold of $r = 24.74$ from below. We intentionally choose a value of $r$ close to this bifurcation point to test our ideas rather than the standard choice of $r = 28$ originating from Lorenz (1963).

We choose $x, z$ to be the 'coarse' variables with $y$ being the variable to be eliminated. Denoting $y = G(x,z)$ and following the recipe of the previous section to generate $G$, we would like to solve

$$\sigma(G-x)\frac{\partial G}{\partial x} + (xG - bz)\frac{\partial G}{\partial z} + G + x(z-r) = 0,$$
(13)

where many of the equations in following the procedure are satisfied identically.

If one is successful in doing so, then the closed coarse theory would take the form
$$\dot{x} = \sigma(G(x,z) - x);$$
$$\dot{z} = xG(x,z) - bz.$$
(14)

*3.1.1 Emergence of Memory Dependence in Coarse Response*

The system (14) would be autonomous if the 'function' $G$ were to be single-valued. Now the question arises as to what sort of solutions one should seek for (13) – global solutions, or local ones with the possibility of allowing multiple 'sheets' for every fixed local region in $x-z$ domain. It is the latter option that appears to be optimal as can be seen from the projection of the phase plot of a single trajectory of (12) onto $x-z$ space (Fig. 1). Fig. 1 shows two (almost) overlapping trajectories; the 'fine' trajectory is computed by solving the full Lorenz system; 'coarse' is a prediction of the 2-dof reduced model to be discussed below. The self-intersections of the trajectory in $x-z$ space are present because it is a projection of a trajectory in the full phase space of the fine system. This may be construed as a simple explanation of history-dependence in coarse behavior of physical systems as arising from observation of coarse features (reduction) of an autonomous fine system, the latter's evolution out of a given state depending only on that state. Geometrically, the intersections imply that at a coarse point $(x,z)$ the actual trajectory travels on different local $G$ sheets (solutions viewed as surfaces, Fig. 2) and for accurate modeling one needs to solve for an adequate number of sheets for systems whose trajectories stretch and turn in bounded regions of phase space (periodic or chaotic behavior). Consequently, we attempt to solve (13) *locally* in blocks in $x-z$ space.

*3.1.2 Algorithm and results*

We look for multiple solutions (sheets) in these blocks with each required to satisfy different imposed data at a single point of the $x-z$ block. This is sensible for at least three reasons:

1. there is no natural choice for initial or boundary data for the problem,
2. imposing data at a single point as opposed to along a non-characteristic curve allows greater freedom in determining *continuous* local invariant manifolds (note that (13) has curved characteristics, in fact solutions of (12), that can intersect),
3. determining non-characteristic surfaces for data specification may not be simple in higher dimensions.

The flip side of this device is that, clearly, the solution for a sheet is not unique; however, for imposed data $y_0$ at $(x_0, z_0)$ the graphs of all possible local solutions in the block satisfying the imposed data contain the actual trajectory of the fine system out of $(x_0, y_0, z_0)$ within the block (sec. 2, (c)). We discretize (13) by the Least Squares Finite Element Method (Jiang, 1998) which is well-suited for hyperbolic systems and also applicable to elliptic problems. A finite element interpolation of $G$ is used to pose an algebraic least-squares problem by squaring the residual, adding the square of the imposed constraint as an addition to the residual (or simply eliminating a degree of freedom for imposed data at a node of the coarse mesh of the block), integrating over the local $x-z$ domain to set up an objective function and looking for a minimizer. We solve the resulting nonlinear optimization problem by the Continuous Simulated Annealing method (Press *et al.*, 1999). Any (approximation of an) absolute minimizer is acceptable; however, our numerical experience with the problem shows that there exist local minima and the simulated annealing method is able to avoid these. Fig. 3 shows a cross section at $x = 0_+$ of a part of the collection of pre-computed sheets in the block $0 < x < 4$, $0 < z < 4$ and Fig. 4 shows a select number of sheets in a particular block.

In evolving the coarse (reduced) model, one identifies the closest sheet in the block to the prescribed initial condition, i.e. some $(x_*, y_*, z_*)$ is specified and in the block containing $(x_*, z_*)$ one identifies the closest $G$ sheet by looking for a minimum of $|G(x_*, z_*) - y_*|$. Information on the direction of the fine vector field evaluated at $(x_*, y_*, z_*)$ can (and should) be incorporated in the choice of the appropriate $G$ function. Using this $G$ sheet one travels to the block boundary, say $(x_1, z_1)$. Now using the 3-tuple $(x_1, G(x_1, z_1), z_1)$ in the generator (rhs) of the fine system (12), one generates a dynamically consistent state in an adjoining block and repeats the procedure. Figs. 1, 5 and 6 show that the coarse theory reproduces the reduced trajectory corresponding to the numerically integrated fine trajectory whenever the existence of a sheet containing the state delivered by evolution from the previous block can be ensured. We construe this as a consistency check for the coarse-graining scheme. The condition is ensured for this test by explicitly solving for sheets with imposed data as the delivered state from the previous block (Fig. 6). It should be noted that even in this case, there are an infinite number of local trajectory segments that have been generated by trying to develop a sheet for one trajectory, and hence a reduction of work has been achieved.

As mentioned earlier, the general strategy is to compute a certain number of local invariant manifolds in every coarse block. In the present implementation for this problem, we consider each coarse block individually; if $(x_i, z_i)$ are the coordinates of the $i^{th}$ corner of the two dimensional block, then a set of points is generated as $(x_i, y_0 + k\Delta y, z_i)$, $k = 0, N$, with $y_0$ chosen from an estimate of the region of phase space to be parametrized and $\Delta y$ being a suitable 'spacing' length. A different sheet $G_{ki}$ is now generated corresponding to each element of the above set such that $G_{ki}(x_i, z_i) = y_0 + k\Delta y$. The procedure is now repeated for each of the four corners of the block and the entire set of $4(N+1)$ functions forms the collection of sheets for the coarse block under consideration. The procedure is now repeated for each coarse block.

We divide the coarse space $-24 \leq x \leq 24$, $0 \leq z \leq 48$ into blocks of $4 \times 4$ units and utilize a $6 \times 6$ finite element mesh of bilinear elements in each block. For each corner of a block we use the values $y_0 = -24$, $\Delta y = 1$, and $N = 48$.

Figure 7 and Fig. 8 show comparisons of actual and coarse solutions using the above set of computed sheets. At interblock transfers, merely a closest sheet is chosen to propagate the coarse solution. Given the extreme sensitivity to initial conditions of the Lorenz system for the parameter range chosen, it is not surprising that the solutions diverge; however, the coarse solution does stay in the correct region of phase space showing history-dependent behavior (self-intersections) (Fig. 7b), while reproducing reasonable running time averages, Fig. 9 and Fig. 10. Here, we define the running time average of a function $f$ up to time $T$ as

$$\bar{f}(T) = \frac{1}{T} \int_0^T f(t) dt. \tag{15}$$

Figures 11, 12, and 13 show comparisons of reduced phase plots for trajectories from three different choices of initial conditions. In all of these cases the results for the running time averages are comparable to the comparisons shown in Figs. 9 and 10.

Of course, it is well-known that conventional notions of convergence with respect to time-step size fail for results from numerical integration of the full Lorenz system with standard schemes. In the results presented here, we use the same ODE integration scheme for both the fine and coarse evolution. The computation of the invariant manifolds, however, is not related to the ODE integration scheme in any way.

### 3.2 Nonlinear Oscillators

The fine set of equations for a 4-dof Hamiltonian system (Chorin et al., 2002) is given as

$$\begin{aligned}\dot{x}_1 &= x_2 \\ \dot{x}_3 &= x_4 \\ \dot{x}_2 &= -x_1\left(1+x_3^2\right) \\ \dot{x}_4 &= -x_3\left(1+x_1^2\right).\end{aligned} \tag{16}$$

$x_1, x_2$ are chosen as the retained coarse variables and $x_3, x_4$ are eliminated. The system has a fixed point at $x_1 = x_2 = x_3 = x_4 = 0$ with all eigenvalues of the Jacobian matrix governing the linearization of the system at the fixed point lying on the imaginary axis, in contrast to hyperbolicity in the Lorenz example.

Denoting $x_3 = G_1(x_1, x_2)$ and $x_4 = G_2(x_1, x_2)$, a 2-d PDE system is obtained as follows,

$$\begin{aligned}x_2 \frac{\partial G_1}{\partial x_1} - x_1\left(1+G_1^2\right)\frac{\partial G_1}{\partial x_2} - G_2 &= 0, \\ x_2 \frac{\partial G_2}{\partial x_1} - x_1\left(1+G_1^2\right)\frac{\partial G_2}{\partial x_2} + G_1\left(1+x_1^2\right) &= 0.\end{aligned} \tag{17}$$

This system has to be solved in 2d coarse space to develop each coarse-to-fine map.

The procedure for solving these equations is similar to the Lorenz example. An arbitrarily chosen bounded domain in $x_1 - x_2$ coarse space is divided into square blocks. We now intend to solve for multiple *pairs* of functions in the blocks; in a block, each such pair is required to satisfy imposed data at one arbitrarily chosen point of the domain. Both the equations in (17) are discretized by the Least Squares Finite element method.

In terms of the discrete solutions, the coarse theory takes the form

$$\begin{aligned}\dot{x}_1 &= x_2, \\ \dot{x}_2 &= -x_1\left(1+G_1^2(x_1, x_2; h)\right),\end{aligned} \tag{18}$$

where we include the extra parameter $h$ in the argument list of $G$ to reflect the fact that information regarding the history of the coarse trajectory is required to ensure a correct evaluation. Of course, (14) would also have to be interpreted similarly. In evolving the coarse (reduced) model, one identifies the nearest pair of 'sheets' $G_1$ and $G_2$ corresponding to the prescribed initial condition in the block, i.e. some $(x_1^*, x_2^*, x_3^*, x_4^*)$ is specified and in the block containing $(x_1^*, x_2^*)$, one identifies a pair of sheets $G_1$ and $G_2$ that minimizes the function $(G_1(x_1^*, x_2^*) - x_3^*)^2 + (G_2(x_1^*, x_2^*) - x_4^*)^2$ amongst all possible candidate pairs in the block. Once a minimizing pair is identified, the coarse theory is evolved using this pair to a coarse block boundary. At the boundary, an interblock transfer is performed by a procedure similar to the one explained for the Lorenz example.

Clearly, the issue of an appropriate phase space metric arises in the identification of the 'nearest' pair of sheets to a given initial condition in a block, in the event that the initial condition does not happen to belong to the range of some pair of sheets for the block. We leave this matter for future examination.

In the example problems to be discussed, the coarse domain was restricted to $-2 < x_1 < 2$, $-2 < x_2 < 2$ and divided into blocks. A minimal set of 'sheet pairs' were calculated in each block each corresponding to a piece of imposed data; a series of points for imposing data, indexed by $k$, is generated by picking a single corner of a block, say $(x_1^c, x_2^c)$, holding $x_3^k$ and $x_4^k$ fixed in turn and varying the other within an arbitrarily chosen range $(-1 \leq x_{3,4}^k \leq 1)$.

Two problems with different initial conditions are considered. Figs.14, 15, 16 and 17 display the performance of the coarse theory. Notice the multiple self-intersection of the phase-plot in the reduced variables indicating history dependence in reduced behavior. There is divergence from the correct solution for the coarse variables simply due to the use of the meager set of pre-computed manifolds. Even so, reasonable match in the time-averaged result for the second example can be observed.

### 3.3 Linear Oscillator

In this section we consider a trivial equation from the dynamics point of view which is non-trivial in the context of arbitrary reduction. The fine dynamical system corresponding to a simple linear spring-mass assembly is given by

$$\dot{x} = -y$$
$$\dot{y} = x. \qquad (19)$$

We retain $x$ as the coarse variable and denote $y = G(x)$ to obtain

$$\frac{dG}{dx} G + x = 0. \qquad (20)$$

As usual, we would like to obtain multiple solutions to this equation in local blocks in $x$ space and subsequently use the coarse model

$$\dot{x} = -G(x; h). \qquad (21)$$

For imposed data $G(x_0) = y_0$, the solution to (20) is

$$G(x) = \pm(r_o^2 - x^2)^{1/2} \, ; \, x_o^2 + y_o^2 = r_o^2. \tag{22}$$

Clearly, a real-valued solution does not exist for all $x$, for specified data $y_0$. Moreover, this situation would be encountered in our solution algorithm even though we seek local solutions, since the coarse local domain is chosen arbitrarily and so is the imposed data. We deal with the situation by seeking complex-valued solutions to (20) in blocks and 'pruning' (ignoring) the obtained sheets at coarse points where a non-zero complex part exists. Fig. 18 shows a few manifolds computed over the chosen coarse domain $-3 \leq x \leq 3$. In this case some sheets for a given block 'end' within the block and not necessarily at the block boundaries, i.e. the sheet is not defined for the entire coarse domain consisting the block. At such sheet boundaries one uses the vector field of the actual fine equation to move to another sheet, either within the block or in another block. The procedure generalizes to higher dimensions.

Figures 19 and 20 show results for trajectories out of two representative initial conditions with the pre-computed set of local invariant manifolds computed with the above restrictions.

Of course, it is an interesting question as to whether such situations would be encountered in the process of exactly solving for the sheets in the examples of the previous sections. There we sought only real-valued discrete solutions (and were able to obtain them numerically). The conservative, but expensive, option in this regard would be to solve for complex-valued functions in all cases; however, some analytical insight leading to a practical tool to help identify the occurrence of such situations, and the degree of their prevalence in general, would be desirable.

### 3.4 A preliminary application to PDE systems: 1-D Elastodynamics of a Strongly Heterogeneous Medium

In this model problem we attempt the spatial averaging of a PDE system. Spatial averaging refers to using a coarse field that is a running spatial average of a fine field over a fixed length scale, say $2\varepsilon$. The typical coarse field corresponding to a fine field $f$ is defined as:

$$\bar{f}(y,t) = \frac{1}{2\varepsilon} \int_{y-\varepsilon}^{y+\varepsilon} f(x,t) dx. \tag{23}$$

We study the model problem of 1-D elastodynamics of a strongly heterogeneous medium. In this model problem, we first simplify the infinite dimensional problem to a set of ODEs such that the PLIM method can be applied to the latter. Then we discuss an approach of dividing the entire domain of interest into sub-domains and the application of the PLIM method to derive the coarse dynamics of a single sub-domain subjected to homogeneous or periodic boundary conditions on the displacement field or to constant acceleration boundary conditions at the ends of the spatial sub-domain. This is followed by a discussion of an attempt to couple such sub-domains to obtain a solution over larger domains of interest.

The equations representing 1D elastic wave propagation are as follows:

$$\left.\begin{aligned}
&\dot{u}(x,t) = v(x,t) \\
&\rho \dot{v}(x,t) = \left(E(x).u_x(x,t)\right)_x \\
&I.C: \quad u(x,0) = \hat{u}(x), \\
&\qquad v(x,0) = \hat{v}(x) \; and \\
&B.C: \quad u(0,t) = u_o(t), u(L,t) = u_L(t) \;\; or \;\; v(0,t) = v_o(t), v(L,t) = v_L(t)
\end{aligned}\right\} 0 < x < L. \quad (24)$$

The fine boundary conditions can, of course, be more general.

Here $u$ and $v$ are the displacement and velocity fields of an elastic medium of length $L$, $\hat{u}, \hat{v}$ are the initial displacement and velocity respectively; and $u_o, u_L$ are the boundary conditions which are restricted to be homogeneous or periodic or developed from the specification of a constant acceleration at the ends. The spatial variation of the Young's modulus $E(\cdot)$ is given as follows,

$$E(x) = E_0 \left(2 + \cos\left(\frac{2\pi x}{\lambda_E}\right)\right) \quad (25)$$

where $\lambda_E$ is the wavelength of the sinusoidal function considered.

*3.4.1 Conversion to system of ODEs*

To apply the method of PLIM to this problem, we first simplify (24) to a system of ODEs by discretizing the fine displacement and velocity as

$$u(x,t) = \sum_{i=1}^{\eta} u^i(t) \varphi^i(x), \quad v(x,t) = \sum_{i=1}^{\eta} v^i(t) \varphi^i(x), \quad (26)$$

where $u^i, v^i$ are the nodal values of $u, v$ at $x^i$ and $\varphi^i(x)$ are the trial functions. Here $\eta$ is the number of discrete nodes over the entire domain. Using this discretization and the Galerkin method of weighted residual for (24), we obtain,

$$\begin{aligned}
&\dot{u}^i = v^i \\
&\dot{v}^i = \beta^{ij} u^j \\
&\text{I.C: } u^i(0) = \hat{u}(x_i), v^i(0) = \hat{v}(x_i) \quad and \\
&\text{B.C: } u^1(t) = u_o(t), u^\eta(t) = u_L(t) \text{ or } v^1(t) = v_o(t), v^\eta(t) = v_L(t) \\
&where \\
&\beta^{ij} = M_{ik}^{-1} \cdot K_{kj}, \\
&M_{ij} = \rho \int \varphi^i(x) \varphi^j(x) dx, \quad K_{ij} = -\int E(x) \varphi^i_{,x}(x) \varphi^j_{,x}(x) dx
\end{aligned} \Bigg\} i,j = 1 \, to \, \eta \qquad (27)$$

This is a coupled system of ODEs in $2\eta$ unknowns, $u^i, v^i$.

Note that in the equation above and in the rest of this section the **summation convention for repeated indices applies.**

*3.4.2 Division of entire domain into sub-domains:*

This model problem is computationally large compared to the other model problems in this paper. For this problem we have to compute sets of $2\eta$ functions that represent the locally invariant manifolds, corresponding to displacement and velocity of each discrete node. To study the application of PLIM to obtain averaged response for this problem we first divide the entire domain $L$ into smaller sub-domains. To begin, we develop a coarse model over one such sub-domain.

Fig. 21 shows a schematic for division of the entire domain into sub-domains and the corresponding change in variation of Young's modulus. Here the entire domain of length $L$ is divided into 4 sub-domains denoted by $sd$. The variation of the Young's modulus over the entire domain has a wavelength $\lambda_E$, thus $L = 8\lambda_E$. A typical sub-domain is of specified length $2\varepsilon = l = 2\lambda_E$ for variation of the Young's modulus (25).

*3.4.3 Implementation of the PLIM method*

On a chosen sub-domain the coarse variables are defined as,

$$\bar{u}(x_o, t) = \frac{1}{2\varepsilon} \int_{x_o - \varepsilon}^{x_o + \varepsilon} u(y, t) \, dy, \quad \bar{v}(x_o, t) = \frac{1}{2\varepsilon} \int_{x_o - \varepsilon}^{x_o + \varepsilon} v(y, t) \, dy \qquad (28)$$

representing the spatial averages of displacement and velocity, over the sub-domain of length $2\varepsilon$, centered about $x_o$. Using the discretization (26) over the sub-domain, the discrete coarse variables may be expressed as,

$$\left.\begin{aligned}\bar{u}(t) &= \tfrac{1}{2\varepsilon}\int_{-\varepsilon}^{\varepsilon} u_i(t)\varphi_i(y)dy = \psi_i\, u_i(t)\\ \bar{v}(t) &= \tfrac{1}{2\varepsilon}\int_{-\varepsilon}^{\varepsilon} v_i(t)\varphi_i(y)dy = \psi_i\, v_i(t)\end{aligned}\right\} i = 1\, to\, \eta \quad (29)$$

where $\psi$ is a time-independent vector of coefficients and now $\eta$ is the number of discrete nodes over the sub-domain. Differentiation of (29) gives the discrete evolution equations for the coarse variables as follows,

$$\left.\begin{aligned}\dot{\bar{u}}(t) &= \psi_i\, \dot{u}_i(t) = \psi_i\, v_i(t)\\ \dot{\bar{v}}(t) &= \psi_i\, \dot{v}_i(t) = \psi_i\, \beta_{im}u_m(t)\end{aligned}\right\} i,m = 1\, to\, \eta. \quad (30)$$

The fine degrees of freedom in (27) are denoted as $u_k = G_k(\bar{u},\bar{v})$ and $v_k = G_{k+\eta}(\bar{u},\bar{v})$, where $u_k, v_k$ are the displacement and velocity of the $k^{th}$ discrete node in the sub-domain. For $\eta$ number of discretized nodes we solve $2\eta$ coupled equations,

$$\left.\begin{aligned}\frac{\partial G_k}{\partial \bar{u}}(\psi_l\, G_{l+\eta}) + \frac{\partial G_k}{\partial \bar{v}}(\psi_l\, \beta_{lm}G_m) &= G_{k+\eta}\\ \frac{\partial G_{k+\eta}}{\partial \bar{u}}(\psi_l\, G_{l+\eta}) + \frac{\partial G_{k+\eta}}{\partial \bar{v}}(\psi_l\, \beta_{lm}G_m) &= \beta_{ki}G_i\end{aligned}\right\} i,k,l,m = 1\, to\, \eta \quad (31)$$

to obtain a set of $2\eta$ functions $G_k$, that represent the locally invariant manifold. We solve the system (31) by discretizing it using Least Squares Finite Element Method and a procedure of Explicit Integration in the direction of time-like coarse variable, detailed in Sawant (2005). The boundary conditions (27)$_{5,6,7,8}$ for the sub-domains are also restricted to homogeneous or periodic or constant acceleration cases, and a collection of sets of functions $G_k$ is computed for different initial conditions (27)$_{3,4}$.

Using the collection of manifolds computed from (31), we can write a closed coarse theory for the coarse variables $\bar{u}, \bar{v}$ as,

$$\left.\begin{aligned}\dot{\bar{u}}(t) &= \psi_i\, G_{i+\eta}(\bar{u}(t),\bar{v}(t))\\ \dot{\bar{v}}(t) &= \psi_i\, \beta_{im}G_m(\bar{u}(t),\bar{v}(t))\end{aligned}\right\} i,m = 1\, to\, \eta. \quad (32)$$

It is important to mention that as the functions $G$ are pre-computed, the obtained coarse theory (32) allows a significant reduction in the computations of the coarse response, than those required to obtain the coarse response from the fine system (27) and (29).

*3.4.4 Results: coarse model for sub-domain*

Fig 22 shows an example where a finite collection of locally invariant manifolds is used to compute the coarse response by evolving (32). The initial displacement is selected as $\hat{u}(y) = 0$ and the initial velocity is chosen as sinusoidal function $\hat{v}(y) = \sin(3\pi y/\lambda)$ with wavelength $\lambda$. The variation of the Young's modulus $E(\cdot)$ is selected as $E_o(2+\sin(2\pi y/\lambda_E))$. The sub-domain is subjected to acceleration driven boundary condition $(i.e.\ \dot{v}^1,\, \dot{v}^\eta = \text{constant})$. In this example the ratio $\lambda/\lambda_E$ is $4/3$.

Figure 22(a) shows the normalized initial conditions imposed on the sub-domain along with the normalized variation of the Young's modulus $E(\cdot)$ and its average obtained from (23), which corresponds to a homogeneous medium and is denoted as $\bar{E}$. For this model problem, we have chosen 20 nodes per wavelength of the variation $E(\cdot)$, as shown in the figure.

Fig 22(b) shows 3 coarse trajectories in the coarse phase space $\bar{u} - \bar{v}$. The actual and coarse trajectories represent the solutions obtained by evolving the fine (27) and coarse (32) systems respectively. The third trajectory represents the solution obtained for a homogeneous medium with Young's modulus $\bar{E}$. Fig. 22(c) and (d) show the evolution of the averaged $\bar{u}, \bar{v}$ in time. Fig. 22 shows that the coarse response obtained by evolving the coarse theory is in good agreement with the actual response and is better than the one for homogeneous medium.

Similar comparisons of the actual and coarse responses corresponding to different initial conditions and variations of Young's modulus also show good agreement up to times equal to 10-20 times the period of the exact solution. However, the solution diverges from the actual response after this time, due to accumulation of errors resulting form the utilization of an inadequate collection of locally invariant manifolds.

In principle, this situation can be improved if we compute many more manifolds corresponding to various initial conditions as imposed data. Deriving an estimate of the required number of pre-computed locally invariant manifolds for sufficient accuracy awaits further study.

*3.4.5 Coupling of sub-domains:*

With the computed PLIMs for a sub-domain in hand, an attempt is made to compute the coarse response for a larger domain formed by coupling the sub-domains. In Fig. 21 the entire domain of length $L$ was divided in 4 sub-domains of length $l = 2\varepsilon = L/4$. In Fig. 23 a domain of length $L = 16\lambda_E$ (normalized to one) is formed by coupling 16 sub-domains of length $l = 2\varepsilon = \lambda_E$.

For this coupling we use a sub-domain with the variation of Young's modulus as $E(y) = E_o(2 + \cos(2\pi y/\lambda_E))$ over $(0 \leq y \leq l)$. Thus the coupled domain has a variation of $E(x) = E_o(2 + \cos(2\pi x/\lambda_E))$ over $(0 \leq x \leq L)$. The normalized initial conditions (24)$_{3,4}$ on the entire domains are $\hat{u} = 0, \hat{v} = \sin(8\pi x)(0 \leq x \leq 1)$, which are shown in Fig. 23. The Figure also shows the Young's modulus $\bar{E}$ of the corresponding homogeneous medium as in Section 4.3.3.

To obtain a space averaged displacement and velocity over the coupled domain, the wave equation (24) is averaged using (23) to obtain

$$\dot{\bar{u}}(x,t) = \bar{v}(x,t), \quad \rho\dot{\bar{v}}(x,t) = (\bar{\sigma}(x))_x,$$
$$\text{I.C:} \quad \bar{u}(x,0) = \hat{\bar{u}}(x), \bar{v}(x,0) = \hat{\bar{v}}(x) \text{ and} \qquad \qquad 0 < x < L \quad (33)$$
$$\text{B.C:} \quad \bar{u}(0,t) = \bar{u}_o(t), \bar{u}(L,t) = \bar{u}_L(t) \text{ or } \bar{v}(0,t) = \bar{v}_o(t),, \bar{v}(L,t) = \bar{v}_L(t)$$

where $\bar{\sigma}$ is the averaged stress expressed, in terms of the fine displacements and the Young's modulus variation in the domain $(x-\varepsilon, x+\varepsilon)$, via (23) as

$$\bar{\sigma}(x,t) = \frac{1}{2\varepsilon} \int_{x-\varepsilon}^{x+\varepsilon} E(y).u_y(y) dy. \qquad (34)$$

The initial conditions $\hat{\bar{u}}, \hat{\bar{v}}$, and the boundary conditions $\bar{u}_o, \bar{v}_o, \bar{u}_L, \bar{v}_L$ are computed from the fine initial conditions (seen in Fig. 23) using (28). The domain as shown in the Fig 23 is divided into 8 coarse quadratic elements. Each coarse element includes 2 sub-domains.

The coarse response $\bar{u}, \bar{v}$ over the entire domain is computed by numerically integrating the Galerkin weak form of the system (33) with 17 coarse nodes. $(\eta = 17)$, which can be re-written as

$$\dot{\bar{u}}^i = \bar{v}^i, \quad \dot{\bar{v}}^i = \alpha^i,$$
$$\text{I.C:} \quad \bar{u}^i(0) = \hat{\bar{u}}(x_i), \bar{v}^i(0) = \hat{\bar{v}}(x_i) \quad \text{and}$$
$$\text{B.C:} \quad \bar{u}^1(t) = \bar{u}_o(t), \bar{u}^\eta(t) = \bar{u}_L(t) \text{ or } \bar{v}^1(t) = \bar{v}_o(t), \bar{v}^\eta(t) = \bar{v}_L(t)$$
$$\alpha^i = M_{ik}^{-1} F_k, \qquad \qquad \qquad \qquad i, j = 1 \text{ to } \eta. \quad (35)$$
$$F_k = -\int \bar{\sigma}(x) \varphi_{,x}^k(x) dx,$$
$$\bar{\sigma}(x,t) = u^i(t) \left[ \frac{1}{2\varepsilon} \int_{x-\varepsilon}^{x+\varepsilon} E(y) \varphi_{,y}^i(y) dy \right],$$

While computing the coarse response it is assumed that the each sub-domain is centered about the gauss points of the coarse mesh of 16 elements. Clearly we know that $\eta = 17$ is a really poor number of coarse nodes to capture the heterogeneity shown in Fig. 23 with a discretization for the fine theory, e.g. (27). However, with the PLIM method we can compute the coarse response, comparable to the actual response obtained by evolving the fine system with 20 nodes per wavelength of $E(\cdot)$ $(i.e. \eta = 321)$. This is achieved by retrieving a good approximation for $\bar{\sigma}$ in (35), at the gauss points at any coarse time step. The utilized algorithm for selecting an appropriate locally invariant manifold follows.

*Selection of PLIM at coarse Gauss points*

The algorithmic goal is to evolve the current coarse state at time $t_n$ with some explicit time-stepping of (35) with information on the current $\bar{u}, \bar{v}$ at a typical Gauss point of the coarse mesh being available from the discrete current coarse fields. Information from the previous time step on the fine state, the averaged variables $\bar{u}, \bar{v}$, and their derivatives are the other important parameters that aid in selection of appropriate manifold at the Gauss point. The following is an algorithmic description for selecting a manifold:

1. Boundary conditions of the sub-domain:
    a. This is achieved from the $\bar{u}, \bar{v}$ and $\bar{u}_x, \bar{v}_x$ at the current time step, known from the coarse evolution (35)
    b. The displacements and velocities at the ends of the sub-domain of length $(l = 2\varepsilon)$ denoted as $u_o, v_o, u_l, v_l$ are approximated as,
    $$u_o = \bar{u} - \bar{u}_x \varepsilon, \quad v_o = \bar{v} - \bar{v}_x \varepsilon, \\ u_l = \bar{u} + \bar{u}_x \varepsilon, \quad v_l = \bar{v} + \bar{v}_x \varepsilon \tag{36}$$
    c. Using the fine states stored from the previous time step, we compute the accelerations at the ends by
    $$a_o = (v_o - v_o^{n-1})/\Delta t, \quad a_l = (v_l - v_l^{n-1})/\Delta t \tag{37}$$
    where $n-1$ refers to the time instant at the beginning of the time interval $[t_{n-1}, t_n]$ and $\Delta t$ is the coarse time step (*which is larger than the fine time step used to solve fine system*)
   
   Fig. 24 shows the schematic of the procedure described above. Here the coarse displacements and velocities and their derivatives are denoted as general variables $\bar{f}, \bar{f}_x$. The figure shows the placement of the sub-domains at Gauss points in the coarse element, the approximation of the end fine variables for the sub-domains from the coarse variables at the Gauss points and a formula to obtain the boundary conditions for the sub-domain.
2. With the end-point accelerations available from Step 1) above and information on the fine state over the subdomain at the end of the previous step, we retrieve a precomputed set of functions representing a locally invariant manifold.
3. Using this locally invariant manifold, an estimate of the current fine states is developed using the current values of $\bar{u}, \bar{v}$ (i.e. $u_k = G_k(\bar{u}, \bar{v})$ and $v_k = G_{k+\eta}(\bar{u}, \bar{v})$). It is important to note that due to the availability of the coarse state at the current time $t_n$ and the assumption that evolution is along a fixed manifold within the coarse time interval $[t_{n-1}, t_n]$, no fine evolution is required to estimate the fine state at $t_n$.

4. Using the fine state we can compute $\bar{\sigma}$ at the current time $t_n$ and compute the *rhs* of $(35)_2$.

*3.4.6 Results: coarse model over the entire domain*

Figure 25 shows the comparison of coarse responses $\bar{u}, \bar{v}$ over the entire domain. The actual and coarse responses are the solutions obtained by evolving the fine and coarse theory respectively. The third response corresponds to the solution obtained over a homogeneous medium. Fig. 25 (a) and (b) show the evolution of displacement and velocity of a coarse node, up to duration equal to 6 times the period of the actual solution. Figures 25 (c) and (d) show the comparison of displacement and velocity over the entire domain at different time instants. All the plots in Fig. 25 show that the coarse and actual solutions are in good agreement for the duration of the time history considered, and that the solution for the homogeneous medium is out of phase from the exact solution.

For this example and some others with homogeneous boundary conditions it was possible to work with much larger time steps for coarse evolution than that required for evolving the fine system. Though the coarse response was limited to short time evolution, the computational effort was also reduced by a factor of 50. The following table gives a summary of computation counts and the efficiency attained by using the method of PLIM for the example shown in Fig. 25. The table gives the count of computations, wall-clock time and the number of time steps used to evolve the fine and coarse system.

| Approach | Computation count | Time | Time steps |
|---|---|---|---|
| Fine | 8457533127 | 17 mins | 80000 |
| Coarse | $166847688 \approx 1/50 \times \text{Fine}$ | 2 mins | $4000 \approx 1/20 \times \text{Fine}$ |

*3.4.7 Remarks on applications to PDE systems*

The preliminary investigation of the application of PLIM to a PDE system reported in this section indicates that 'filtering', i.e. running space averaging, followed by a finite dimensional approximation provides a systematic means of a first reduction of dimensionality of the fine model that has to be coarse grained. We show modest success, especially considering this to be a first implementation of our ideas to this class of problems, and it is clear that the approach can benefit from advances at a theoretical level that provide more understanding of the algorithm. It is our belief that utilizing coarse variables as space-*time* averaged fields would greatly reduce the number of locally invariant manifolds required for good coarse model accuracy as well as provide significant savings on time-stepping. Time-averaging of ODE systems using the PLIM methodology is discussed in Acharya and Sawant (2005), and these ideas can be applied to space-time averaged PDE discretization along the lines described in this Section. It would also be of interest to see how well these ideas can be used in the closure step in the setting of the Variational Multiscale Method of Hughes and co-workers (2005).

## *4. Concluding remarks*

We have demonstrated a scheme for model reduction of nonlinear ODE systems. The scheme requires minimal hypotheses. A by-product is the approximate reconstruction of solutions of the actual system being reduced.

The idea appears to shed light on physical behavior. In particular, physically observed history-dependence in coarse observation of fine scale dynamics can be interpreted, and described, in terms of our scheme. In the context of homogenization of hyperbolic PDE, Tartar (1990) and Amirat *et al*. (1992) show the emergence of memory effects in a deterministic context. The emergence of non-Markovian behavior in coarse response is established in (Chorin et al., 2000, 2002) based on concepts of the Mori-Zwanzig Projection Operator Theory in statistical mechanics. This theory also indicates that a full knowledge of fine initial conditions is required, in general, for an exact reduction to a small number of coarse variables. Interestingly, our approach indicates the same requirement as well as indicating the emergence of memory, but based on completely different logical arguments.

The Projection Operator technique begins by replacing the non-closed, right hand-side of the evolution statement for the coarse variables by a Markovian approximation plus the difference between the exact right hand side and the Markovian term. The difference is shown to be represented formally exactly by a memory term in the coarse variables and a noise term, the noise function being determined as the solution of a system of hyperbolic linear partial differential equations (e.g. Givon et al., 2004). First-order Optimal Prediction Methods (Chorin et al., 2000) work with a reduced dynamics that includes only the Markovian approximation and Optimal Prediction with Memory (Chorin et al., 2002) includes an approximation of the memory term.

In comparing our scheme, that also relies on solving a PDE, with the Projection Operator Technique, we first note that for most nonlinear vector fields of the fine dynamical system involved, analytical solutions to the PDE in both approaches are unlikely. Of course, in the Projection Operator Technique the system of PDEs is linear whereas in our case it is quasilinear, but for highly oscillatory vector fields solving the linear hyperbolic system is not a simple matter. It is perhaps fair to say that for the highly oscillatory case, solutions may only be sought for weak limits. However these 'averaged' solutions are not unique with imposed initial data only on the weak limits as shown in Menon (2002) for the case of non-divergence-free fine vector fields (e.g. dissipative fine dynamics). Nevertheless, the linear structure of the Projection Operator equations does enable formal perturbative approximations. Computationally, the domain of the defining mapping to be solved for in our scheme is the coarse-phase space (low-dimensional), its range being the high-dimensional phase space of the fine dynamics, whereas for the noise function in the Projection Operator Technique the requirement is exactly the opposite along with the addition of time as an extra independent variable. Given these stipulations, it is well understood that the computational burden for our equations, assuming both sets of PDE were to be approximated computationally, would be much less severe than in the Projection Operator Technique.

Our scheme points to the natural emergence of stochastic coarse response of a deterministic fine system as arising from two sources:

1. an incomplete knowledge of 'fine' initial conditions;
2. jumps between local invariant manifolds parametrized by coarse variables, at singularities of the defining PDE for these manifolds (discussion surrounding (11)). In our approach here, we have used the fine infinitesimal generator to define these jumps correctly, but if that were not to be available, e.g. if coarse variables are the only 'observables' with no knowledge of fine dynamics being available, then the physical outcome would necessarily have to be interpreted as a random event.

Computationally, our method appears to be suitable for reduction (e.g. subgrid modeling) even in the absence of a separation of time-scales in the evolution of the fine variables; in fact, it seems that it is best suited for such situations and in this sense it is complementary to averaging approaches for slow-fast systems, e. g. Geometric Singular Perturbation Theory (Jones, 1994), Homogenization in time (Bornemann, 1998).

Mathematically, there seems to be connections between our idea and those of the theory of Minimal Foliations (Moser, 1988) and Anti-integrable systems (Aubry *et al*. 1995, 1990, 1983). From the purely practical standpoint of exercising our method robustly, it would be useful to have in hand a constructive existence result for the local invariant manifolds we seek and the maximal domain of existence of such for given data; we hope that such a result would also shed light on when a formulation in complex variables – as in Sec. 3.3 - of the problem is required. It is also natural to expect that such a result would have to address point (d) of Sec. 2. Perhaps

Gromov's h-principle method is relevant in this regard, but assessing its relevance with precision is beyond our mathematical competence at this point in time.

In this paper, the scheme has been applied to the reduction of *small* fine systems displaying non-trivial dynamics. There do not appear to be major conceptual or algorithmic barriers for extensions to moderately large systems. For practical success in the modeling of spatially discretized PDE systems, one would clearly have to utilize the scheme in cells of the size of the coarse theory resolution, with cell boundary conditions treated as parameters (within each time step) because of the applicability of the theory to only autonomous systems. By a standard device, parameters can be treated as coarse variables in our methodology as in Section 3.4. While this treatment of cell boundary conditions may suffice for computational purposes, the restriction to autonomous systems appears to be a major conceptual shortcoming of the methodology. Even with these simplifications, computational implementation for practical systems promises to be an interesting challenge for modern computational science and engineering.

## 5. Acknowledgment

We thank Prof. R. D. Moser for suggesting the Lorenz system as a test of methodology. Financial support for this work was provided by the US ONR (N00014-02-1-0194) and the US AFOSR (F49620-03-1-0254).

**List of Figures:**

Fig.10. L1: Running time averages of absolute values of fine and coarse solutions
 (a) $\overline{|x|}$ vs.t, (b) $\overline{|y|}$ vs.t, (c) $\overline{|z|}$ vs.t ($IC: x=0, y=2, z=8$)

Fig.11. L2: Trajectories in reduced phase space (a) fine, (b) coarse ($IC: x=-10, y=5, z=23$)

Fig.12. L3: Trajectories in reduced phase space (a) fine, (b) coarse ($IC: x=1, y=-5, z=12$)

Fig.13. L4: Trajectories in reduced phase space (a) fine, (b) coarse ($IC: x=10, y=5, z=13$)

Fig.14. H1: Trajectories in reduced phase space (a) fine, (b) coarse
 ($IC: x_1 = -0.875, x_2 = -0.875, x_3 = 0.5, x_4 = 0.5$)

Fig.15. H1: Fine and coarse solutions in time (a) $x_1$ vs.t, (b) $x_2$ vs.t,
 (c) $x_3$ vs.t, (d) $x_4$ vs.t ($IC: x_1 = -0.875, x_2 = -0.875, x_3 = 0.5, x_4 = 0.5$)

Fig.16. H2: Trajectories in reduced phase space (a) fine, (b) coarse
 ($IC: x_1 = 1.125, x_2 = 1.125, x_3 = 0.5, x_4 = 0.5$)

Fig.17. H2: Running time averages of fine and coarse solutions
 (a) $\overline{x_1}$ vs.t, (b) $\overline{x_2}$ vs.t, (c) $\overline{x_3}$ vs.t, (d) $\overline{x_4}$ vs.t ($IC: x_1 = 1.125, x_2 = 1.125, x_3 = 0.5, x_4 = 0.5$)

Fig.18. Some Manifolds in adjacent blocks (Linear Oscillator)

Fig.19. C1: Trajectories in phase space ($IC: Ex1: x=-0.3, y=-1.8; Ex2: x=0.2, y=1.0$)

Fig.20. C2: Fine and coarse solutions in time Ex1: (a) $x$ vs.t, (b) $y$ vs.t ($IC: x=-0.3, y=-1.8$)
 Ex2: (c) $x$ vs.t, (d) $y$ vs.t ($IC: x=0.2, y=1.0$)

Fig.21. Schematic diagram showing the sub-domains

Fig.22. Comparison of exact and coarse response Example 2: $\hat{u}(y)=0, \hat{v}(y)=\sin(3\pi y)$
 (a) Initial conditions and $E(y)$, (b) Coarse Trajectory $\bar{u}$ vs. $\bar{v}$, (c) $\bar{u}$ vs.t, (d) $\bar{v}$ vs.t

Fig.23. Variation of $E(x)$ and initial conditions over a larger domain formed by 16 sub-domains

Fig.24. Selection of PLIM at gauss points

Fig.25. Coarse response over the entire domain Evolution in time: (a) $\bar{u}$ vs. $t$ (b) $\bar{v}$ vs. $t$
 Spatial variation: (c) $\bar{u}$ vs. $x$ (d) $\bar{v}$ vs. $x$

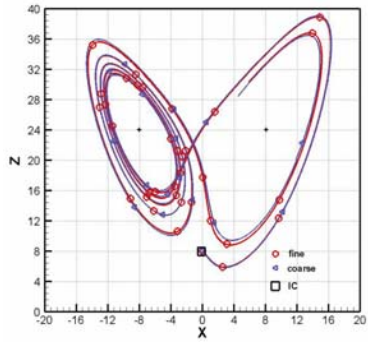

Fig.1. Fine and coarse trajectories in reduced phase space
($IC: x = 0, y = 2, z = 8$)

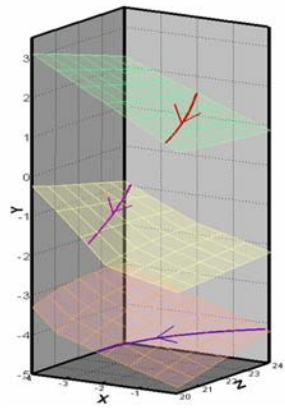

Fig.2. Segments of a single trajectory on multiple invariant sheets

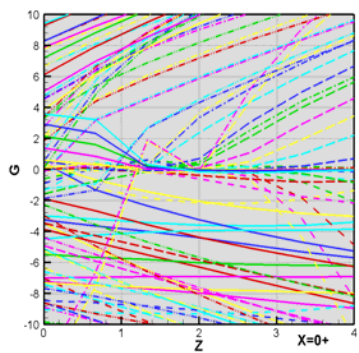

Fig.3. Sections of Manifolds between limits $-10 \leq y \leq 10$ at $x = 0_+$ in the block $0 \leq x \leq 4, 0 \leq z \leq 4$ (Lorenz Example)

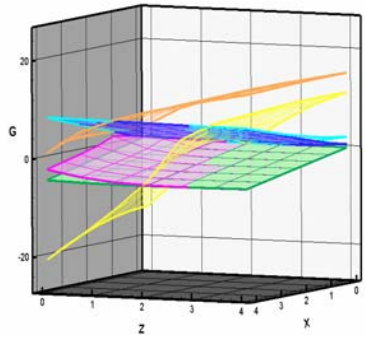

Fig.4. 3-D View of Few Manifolds in a block $(0 \leq x \leq 4, 0 \leq z \leq 4)$

(Lorenz Example)

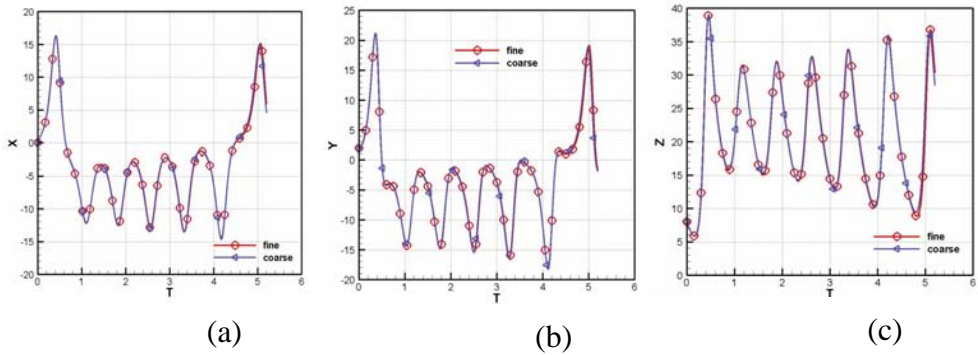

(a)          (b)          (c)

Fig.5. Fine and coarse solutions in time $(a)\, x\ vs.t, (b)\, y\ vs.t, (c)\, z\ vs.t$

($IC: x = 0,\ y = 2,\ z = 8$)

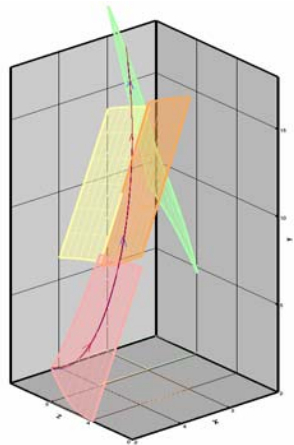

Fig.6. Inter-block transfers

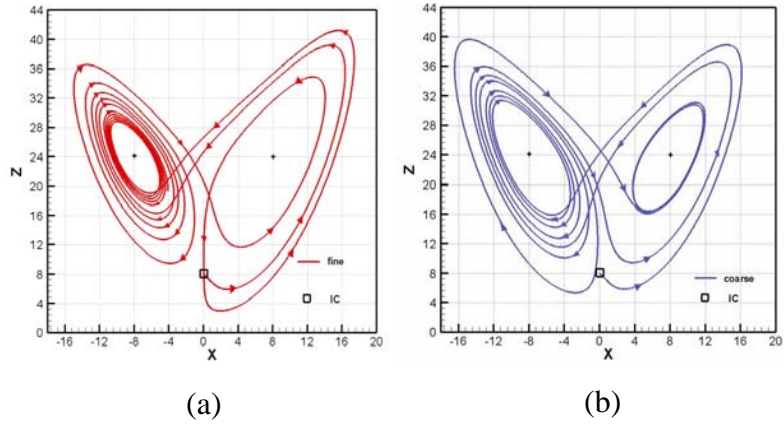

(a)                  (b)

Fig.7. L1: Trajectories in reduced phase space (a) fine, (b) coarse ($IC: x=0, y=2, z=8$)

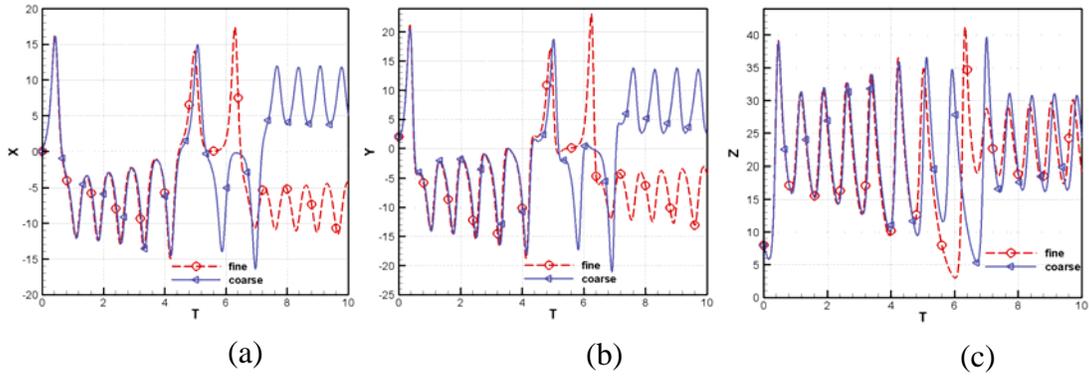

(a)             (b)             (c)

Fig.8. L1: Fine and coarse solutions in time
(a) $x$ vs.$t$, (b) $y$ vs.$t$, (c) $z$ vs.$t$ ($IC: x=0, y=2, z=8$)

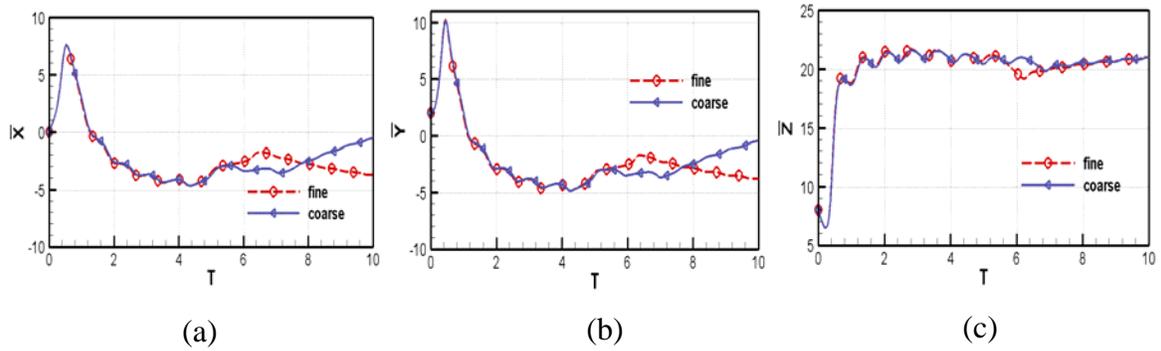

(a)             (b)             (c)

Fig.9. L1: Running time averages of fine and coarse solutions
(a) $\bar{x}$ vs.$t$, (b) $\bar{y}$ vs.$t$, (c) $\bar{z}$ vs.$t$ ($IC: x=0, y=2, z=8$)

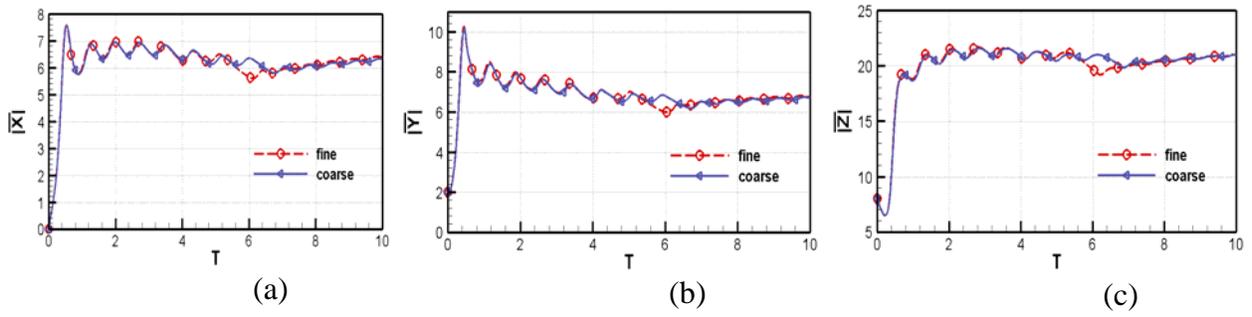

Fig.10. L1: Running time averages of absolute values of fine and coarse solutions $(a)\overline{|x|}\,vs.t,(b)\overline{|y|}\,vs.t,(c)\overline{|z|}\,vs.t$ $(IC: x=0, y=2, z=8)$

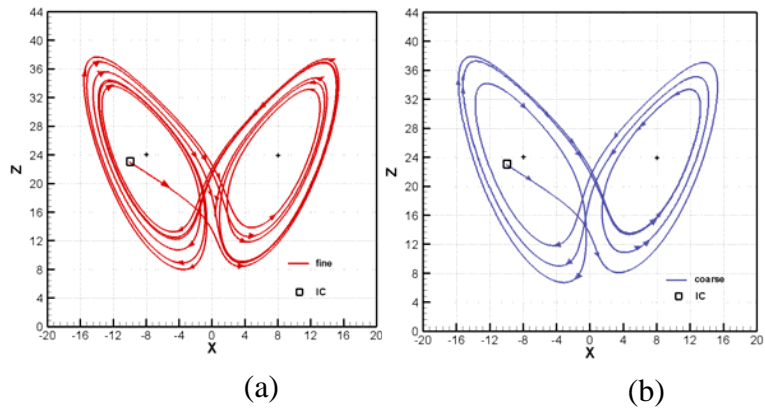

Fig.11. L2: Trajectories in reduced phase space $(a)$ fine, $(b)$ coarse $(IC: x=-10, y=5, z=23)$

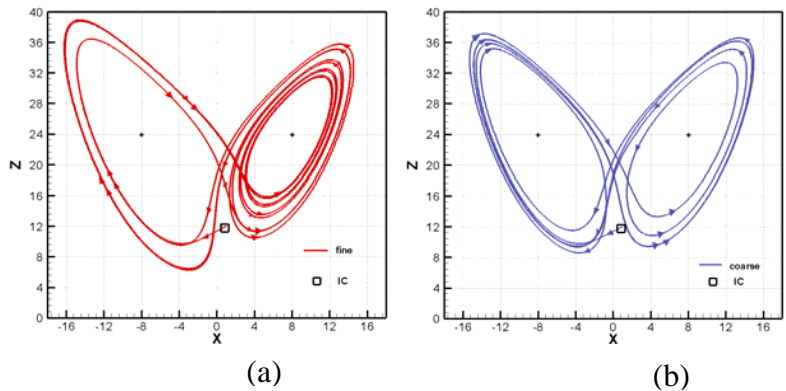

Fig.12. L3: Trajectories in reduced phase space $(a)$ fine, $(b)$ coarse $(IC: x=1, y=-5, z=12)$

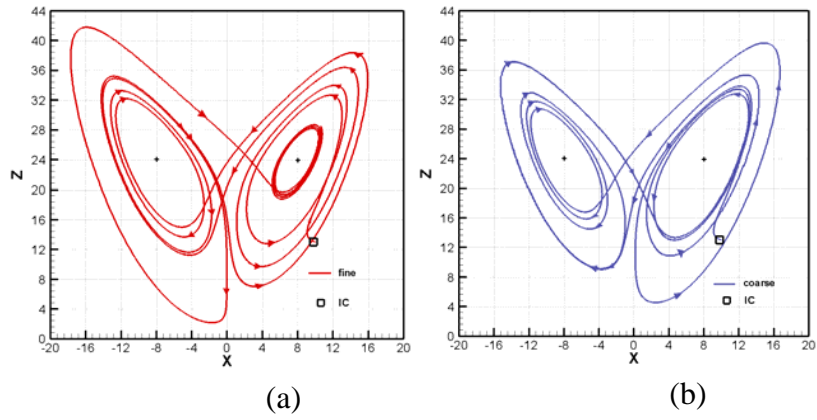

Fig.13. L4: Trajectories in reduced phase space $(a)$ fine, $(b)$ coarse $(IC: x=10, y=5, z=13)$

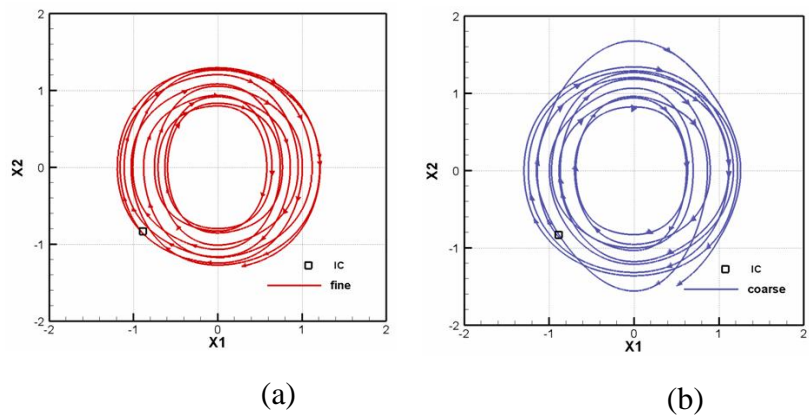

Fig.14.H1: Trajectories in reduced phase space $(a)$ fine, $(b)$ coarse
$(IC: x_1 = -0.875, x_2 = -0.875, x_3 = 0.5, x_4 = 0.5)$

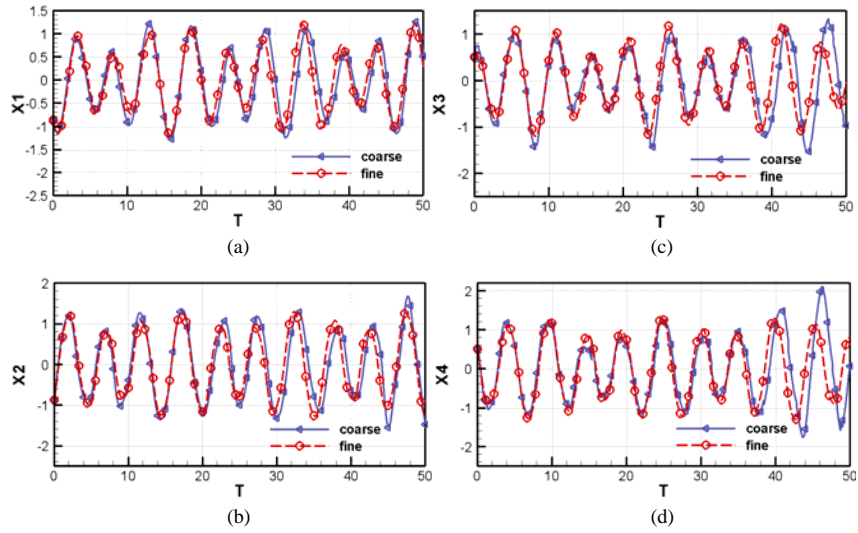

Fig.15. H1: Fine and coarse solutions in time $(a) x_1 \ vs.t, (b) x_2 \ vs.t,$ $(c) x_3 \ vs.t, (d) x_4 \ vs.t$ ($IC: x_1 = -0.875, x_2 = -0.875, x_3 = 0.5, x_4 = 0.5$)

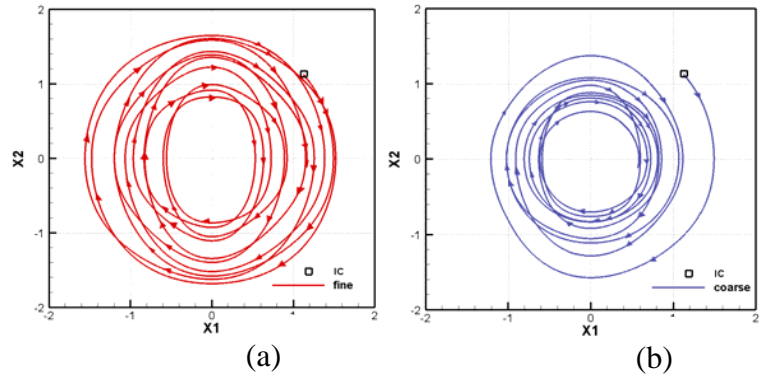

Fig.16. H2: Trajectories in reduced phase space $(a)$ fine, $(b)$ coarse ($IC: x_1 = 1.125, x_2 = 1.125, x_3 = 0.5, x_4 = 0.5$)

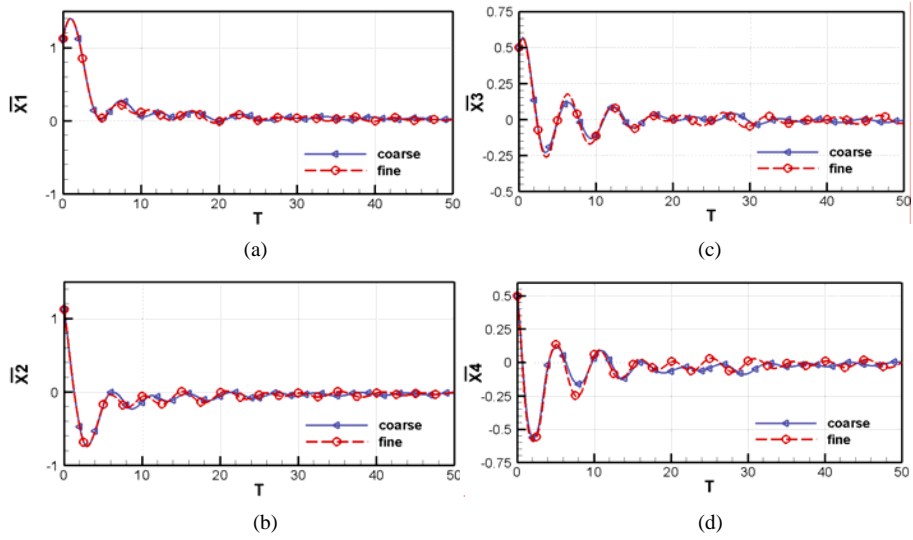

Fig.17. H2: Running time averages of fine and coarse solutions
$(a)\bar{x}_1\ vs.t, (b)\bar{x}_2\ vs.t, (c)\bar{x}_3\ vs.t, (d)\bar{x}_4\ vs.t$ $(IC: x_1 = 1.125, x_2 = 1.125, x_3 = 0.5, x_4 = 0.5)$

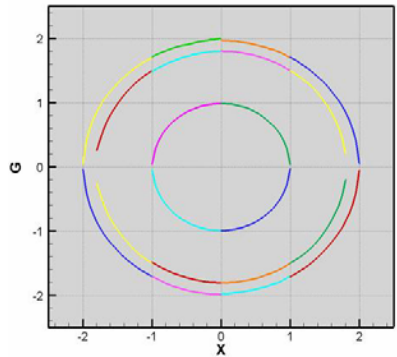

Fig.18. Some Manifolds in adjacent blocks (Linear Oscillator)

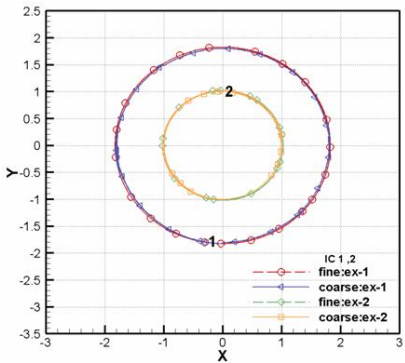

Fig.19. C1: Trajectories in phase space $(IC: Ex1: x = -0.3, y = -1.8;$
$Ex2: x = 0.2, y = 1.0)$

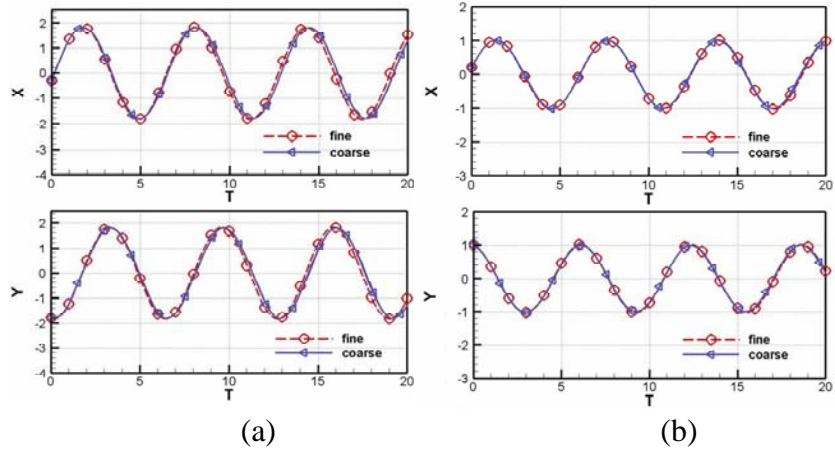

(a)                  (b)

Fig.20. C2: Fine and coarse solutions in time Ex1: $(a)\, x\ vs.t, (b)\, y\ vs.t$ $(IC: x=-0.3, y=-1.8)$

Ex2: $(c)\, x\ vs.t, (d)\, y\ vs.t$ $(IC: x=0.2, y=1.0)$

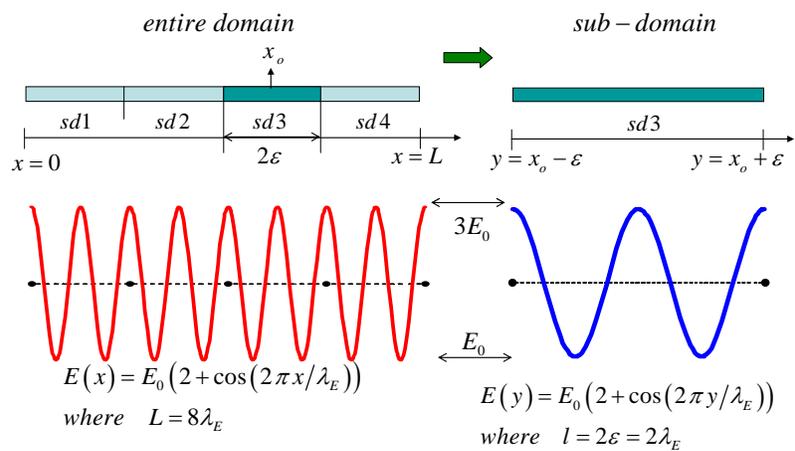

Fig.21. Schematic diagram showing the sub-domains
35

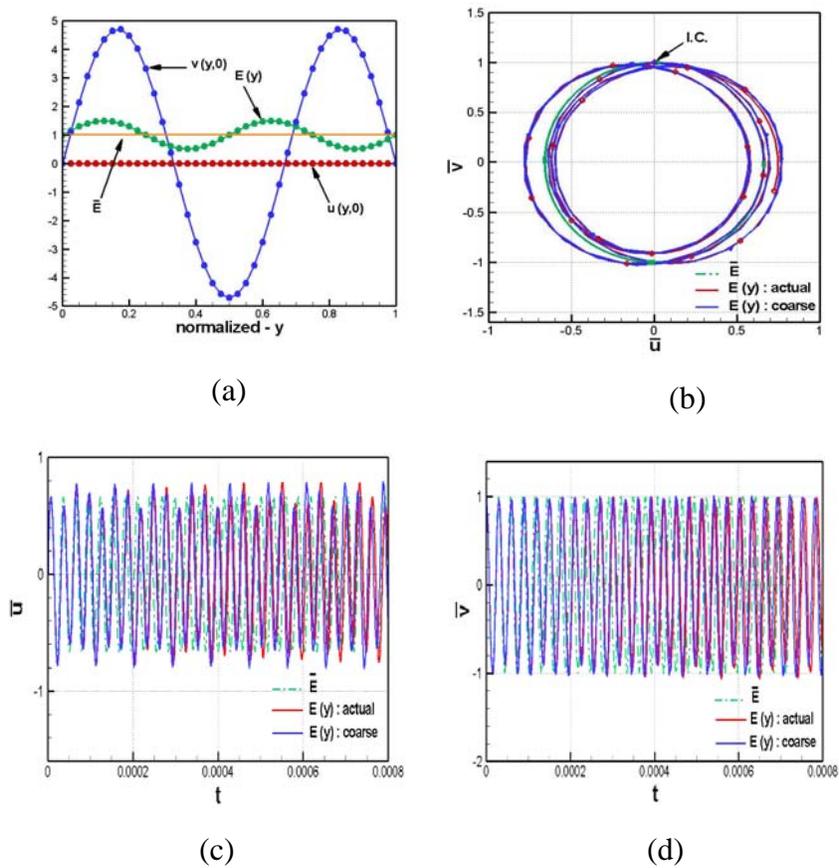

Fig.22. Comparison of exact and coarse response Example 2: $\hat{u}(y)=0, \hat{v}(y)=\sin(3\pi y)$
(a) Initial conditions and $E(y)$, (b) Coarse Trajectory $\bar{u}\ vs.\bar{v}$, (c) $\bar{u}\ vs.t$, (d) $\bar{v}\ vs.t$

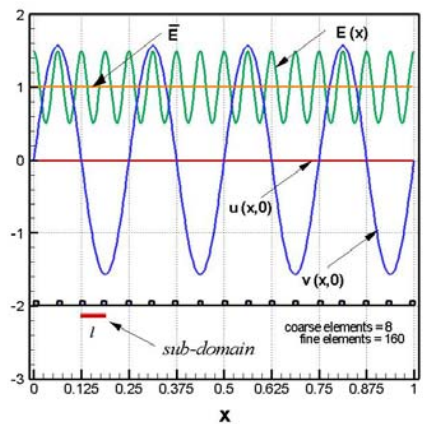

Fig.23. Variation of $E(x)$ and initial conditions over a larger domain

formed by 16 sub-domains

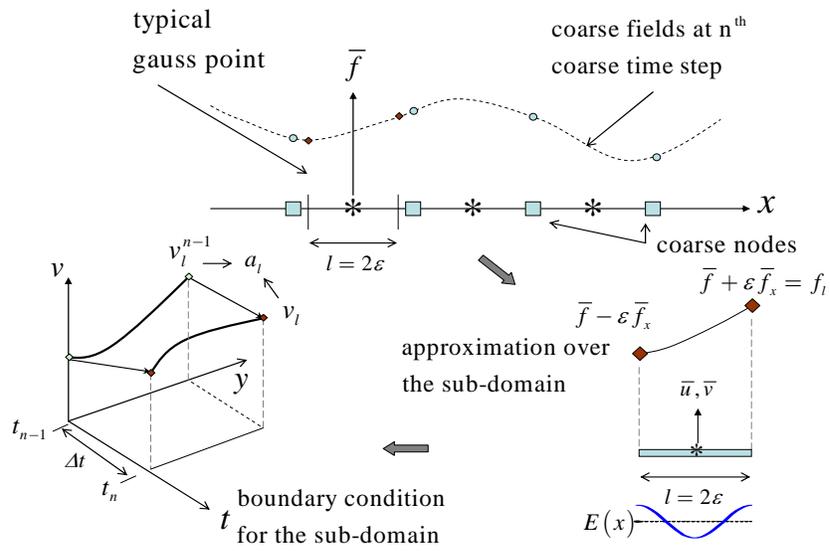

Fig.24. Selection of PLIM at gauss points

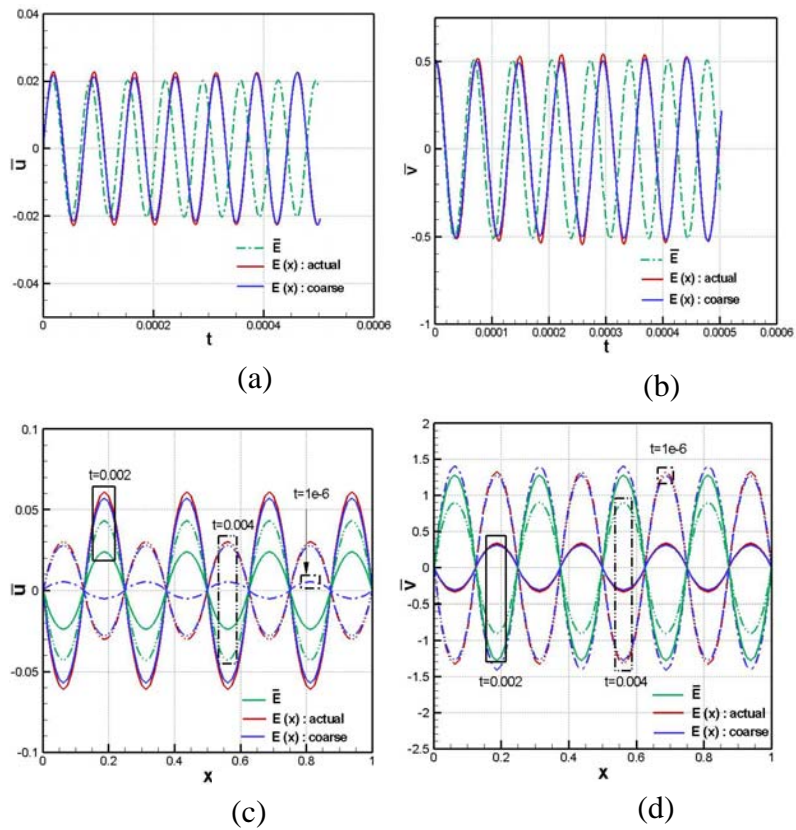

Fig.25. Coarse response over the entire domain Evolution in time: (a) $\bar{u}$ vs. $t$ (b) $\bar{v}$ vs. $t$
Spatial variation: (c) $\bar{u}$ vs. $x$ (d) $\bar{v}$ vs. $x$